\documentclass[12pt]{amsart}

\usepackage{url}
\usepackage{hyperref}
\usepackage[utf8]{inputenc} 

\newtheorem{lemma}{Lemma}
\newtheorem{theorem}[lemma]{Theorem}
\usepackage{graphicx} 

\usepackage{booktabs} 
\usepackage{array} 
\usepackage{paralist} 
\usepackage{verbatim} 
\usepackage{subfig} 
\usepackage{amsmath}
\usepackage{amssymb}
\usepackage{mathtools}

\begin{document}

\title[Beyond the excised ensemble]{Beyond the excised ensemble: modelling elliptic curve $L$-functions with random matrices}

\abstract
The `excised ensemble', a random matrix model for the zeros of quadratic twist families of elliptic curve $L$-functions, was introduced by Due{\~n}ez, Huynh, Keating, Miller and Snaith \cite{kn:dhkms12}. The excised model is motivated by a formula for central values of these $L$-functions in a paper by Kohnen and Zagier \cite{kohnen_zagier}.  This formula indicates that for a finite set of $L$-functions from a family of quadratic twists, the central values are all either zero or are greater than some positive cutoff.  The excised model imposes this same condition on the central values of characteristic polynomials of matrices from $SO(2N)$.  Strangely, the cutoff on the characteristic polynomials that results in a convincing model for the $L$-function zeros is significantly smaller than that which we would obtain by naively transferring Kohnen and Zagier's cutoff to the $SO(2N)$ ensemble.  In this current paper we investigate a modification to the excised model.  It lacks the simplicity of the original excised ensemble, but it serves to explain the reason for the unexpectedly low cutoff in the original excised model.  Additionally, the distribution of central $L$-values is  `choppier' than the distribution of characteristic polynomials, in the sense that it is a superposition of a series of peaks: the characteristic polynomial distribution is a smooth approximation to this.  The excised model didn't attempt to incorporate these successive peaks, only the initial cutoff.  Here we experiment with including some of the structure of the $L$-value distribution.  The conclusion is that a critical feature of a good model is to associate the correct mass to the first peak of the $L$-value distribution. 
\endabstract

\author{I.A. Cooper}
\address{School of Mathematics, University of Bristol, Bristol, BS8 1TW, United Kingdom} \email{ian.cooper@bristol.ac.uk}

\author{Patrick W. Morris} \address{Berlin Mathematical School, TU Berlin, Kekr. MA 2-2, Strasse des 17. Juni 136, Berlin, Germany 10623} \email{patrick.morris@fu-berlin.de}

\author{  N.C. Snaith}\address{School of Mathematics, University of Bristol, Bristol, BS8 1TW, United Kingdom} \email{N.C.Snaith@bris.ac.uk}

\thanks{
The first author is supported on an EPSRC Doctoral Training Account.  The second author was supported by the London Mathematical Society on the Undergraduate Research Bursary Scheme, Ref. URB14/16.}

\maketitle

\section{Introduction}

The use of random matrix theory to model statistics of $L$-functions grew out of the connection between eigenvalue statistics of matrices in the unitary group and the statistics of the zeros of the Riemann zeta function (\cite{kn:mont73,kn:hejhal94,kn:odlyzko89}, or see \cite{zeta} for a review). Katz and Sarnak \cite{kn:katzsarnak99a,kn:katzsarnak99b} suggested that the zero statistics of families of $L$-functions, rather than the zero statistics of individual $L$-functions, could also be modelled by eigenvalue statistics from the classical compact groups. In this paper, we shall look at families of even quadratic twists of elliptic curve $L$-functions. According to the Katz and Sarnak philosophy, in the limit of large conductor (a parameter that orders the $L$-functions in this family), the zero statistics should tend to the statistics of the zeros of random matrix characteristic polynomials from $SO(2N)$. Here the size of $N$ is chosen to equate the density of the eigenvalues on the unit circle with the density of the zeros near the central point (where the critical line crosses the real axis). 

From this, it is reasonable to expect the $L$-function values to be modelled by the characteristic polynomials of $SO(2N)$ matrices. We define the characteristic polynomial of a matrix $B\in SO(2N)$ as
\begin{equation}
\Lambda_B (s)=\prod_{j=1}^N (1-se^{i\theta_j})(1-se^{-i\theta_j})
\end{equation}
where the eigenvalues of $B$ are $e^{\pm i\theta_1},\ldots,e^{\pm i\theta_N}$.  This extension from zero statistics to value statistics of $L$-functions was first investigated by Keating and Snaith \cite{kn:keasna00a}, who used the value distribution and moments of the characteristic polynomial to conjecture the value distribution and moments of families of $L$-functions \cite{values}. The correspondence between values of characteristic polynomials and $L$-function values will be of central importance in this work.

A conjecture for averages of ratios of $L$-functions by Conrey, Farmer and Zirnbauer \cite{kn:cfz2} gives a very accurate expression for the one-level density of zeros in families of $L$-functions. This was tested for the family of even quadratic twists of an elliptic curve $L$-function in \cite{one_level_density}. The Katz-Sarnak philosophy tells us that as we sample $L$-functions of higher and higher conductor in our family, the zero statistics should tend to the statistics of eigenvalues from the special orthogonal group, and the ratios conjecture prediction for the one-level density quantifies how we expect this limit to be approached.

For $L$-functions with relatively small conductor, the one-level density prediction from the ratios conjecture fails to account for a number theoretic phenomenon affecting the distribution of the lowest zeros (those nearest the central point). This was first observed by Miller \cite{kn:mil05} in a different family of $L$-functions associated with elliptic curves. For the case of quadratic twists (with an even functional equation) of an elliptic curve $L$-function, this was discussed in \cite{kn:dhkms12,finite_conductor}, where a modification to the $SO(2N)$ ensemble was proposed (the excised model referred to in the abstract) which recovered the statistics of the lowest zero in the finite conductor regime. 

However, while this model captured the qualitative features of the data well, it was not fully satisfactory in quantifying them, as the most natural model for the zero statistics (detailed in \cite{vanishings}) resulted in an obvious discrepancy between the frequency of vanishing of central values from the family of $L$-functions and the corresponding random matrix statistic. This appeared to be at odds with the interpretation of the characteristic polynomials of the excised ensemble as a smooth approximation to the family of $L$-functions for finite conductor. It is therefore natural to ask if further modifications to the $SO(2N)$ ensemble could further improve on the excised model, which was the motivation for the present work.

The model we propose here is not one we advocate for day-to-day use in modelling $L$-functions, as the original excised model has a simplicity that makes it more appropriate for that.  However, the effectiveness of the modifications described here explain why the excised model gives a reasonable result. This is the main result of this paper. 

\subsection{Families of elliptic curve $L$-functions}\label{sect:elliptic}
We shall consider twists of a fixed elliptic curve $E/ \mathbb{Q}$ by a quadratic character $\chi_{d}$. For an elliptic curve, the associated $L$-function may be written as an Euler product for $s > 1$: 
\begin{equation}
L_E(s) = \prod_{p \nmid M} \left(1 - \frac{\lambda_p}{p^s} \right)^{-1} \prod _{p | M} \left(1 - \frac{\lambda_p}{p^s} + \frac{1}{p^{2s}} \right)^{-1}
\end{equation}
which may be extended via analytic continuation to a meromorphic function on the complex plane. In the above formula, and throughout the paper, $M$ is the conductor of the elliptic curve. Here, the $\lambda_p$s are related to $\#E(\mathbb{F}_p)$, the number of integer points on the curve over the finite field of size $p$:
\begin{equation}
\lambda_p = \frac{p+1 - \#E(\mathbb{F}_p)}{\sqrt{p}}.
\end{equation}
The twist of an elliptic curve $L$-function by a quadratic character $\chi_d$ (with $d$ a fundamental discriminant) is given by changing the Euler product in the definition above to
\begin{equation}
\prod_{p \nmid M} \left(1 - \frac{\lambda_p \chi_d(p)}{p^s} \right)^{-1} \prod _{p | M} \left(1 - \frac{\lambda_p \chi_d(p)}{p^s} + \frac{\chi_d(p)^2}{p^{2s}} \right)^{-1}
\end{equation}
for $s>1$. $L_E(s,\chi_d)$ is the analytic continuation of this product and is the $L$-function associated with a different elliptic curve $E_d$.

When looking at data for a finite number of twists, we will consider the family of all quadratic twists of an elliptic curve $L$-function with $0<d \leq X$ (or $-X\leq d<0$).
The behaviour of $L_E(s, \chi_{d})$ at $s=\frac{1}{2}$ depends on the sign $\chi_d(-M)\omega(E)$ in the functional equation relating $L_E(s,\chi_d)$ to $L_E(1-s,\chi_d)$: 
\begin{equation}
L_E(s,\chi_d) = \chi_d(-M)\omega(E) \left( \frac{2 \pi}{\sqrt{M} |d|} \right)^{2s-1} \frac{\Gamma(3/2-s)}{\Gamma(s + 1/2)} L_E(1-s,\chi_d).
\end{equation}
As such, we will treat twists with even versus odd signs in their functional equation as separate families (referred to as $\mathcal{F}^{+}$ and $\mathcal{F}^{-}$ respectively). It is $\mathcal{F}^{+}$ that we concentrate on in this paper.

We will also define $\mathcal{F}^{0}$ and $\mathcal{F}^{1}$ as subfamilies of  $\mathcal{F}^{+}$ and $\mathcal{F}^{-}$, where the order of vanishing at $s=\frac{1}{2}$ is 0 and 1 respectively. With all these families, the set of fundamental discriminants $d$ such that $L_E(s,\chi_{d}) \in \mathcal{F}^{i}$ will be referred to as $\mathcal{D}^{i}$.  The conjecture of Birch and Swinnerton-Dyer tells us that the order of vanishing at $s=\frac{1}{2}$ is equal to the rank of the Mordell-Weil group of rational points on the elliptic curve (referred to simply as the rank of the elliptic curve).

\section{Review of the excised model} \label{sect:excised}

\subsection{Motivation}
Following the `recipe' described in \cite{kn:cfz2} (for the original case of moments see \cite{kn:cfkrs}) one obtains a conjectural expression for the average over $\mathcal{F}^{+}$ of the ratio $R_E (\alpha,\gamma)$:
\begin{equation}
R_E (\alpha,\gamma) \coloneqq \sum_{\substack{0 < d \leq X \\ d \in \mathcal{D}^+}}  \frac{L_E(1/2 + \alpha,\chi _d)}{L_E(1/2 + \gamma,\chi _d)}.
\end{equation}
Using this ratio conjecture, one can derive a formula for the one-level density of the zeros. The one-level density is defined as
\begin{equation}
S_1(f) = \sum_{\substack{0 < d \leq X \\ d \in \mathcal{D}^+}} \sum_{\gamma_d} f(\gamma_d),
\end{equation}
where the $\gamma_d$ are the heights on the critical line of the zeros of $L_E(s,\chi_d)$ and $f$ is a suitable test function, for example an even Schwarz function. The result, given by Huynh, Keating and Snaith, is:
\begin{theorem}(Theorem 2.3, \cite{one_level_density})  Assuming the Ratios Conjecture,
the 1-level density for the zeros of the family of even quadratic twists of an elliptic curve $L$-function $L_E(s)$ with prime conductor $M$ is given by
\begin{align} \label{oneleveldensity}\nonumber
S_1(f) = &~
\frac{1}{2\pi} \int_{-\infty}^\infty f(t) \sum_{\substack{0<d \leq X\\
d\in \mathcal{D}^+}} \Bigg( 2\log
\left(\frac{\sqrt{M}|d|}{2\pi} \right) + \frac{\Gamma'}{\Gamma}(1
+ it) + \frac{\Gamma'}{\Gamma}(1 - it)\nonumber
\\  & + 2\Big[-\frac{\zeta'(1 + 2it)}{\zeta(1 + 2it)} +
\frac{L_E'({\rm sym}^2, 1 + 2it)}{L_E({\rm sym}^2, 1+2it)} +
A_E^1(it,it)
\\ \nonumber & - \left(\frac{\sqrt{M}|d|}{2\pi}\right)^{-2it} \frac{\Gamma(1 - it)}{\Gamma(1 + it)}
\frac{\zeta(1 + 2it)L_E({\rm sym}^2, 1 -
2it)}{L_E({\rm sym}^2,1)}A_E(-it, it)\Big] \Bigg)dt \\
 & + O(X^{1/2+\varepsilon}),\nonumber
\end{align}
where $f$ is an even test function, $L_E({\rm sym}^2,s)$ is the associated symmetric square $L$-function, and $A_E$ and $A_E^1$ are arithmetic factors given in \cite{one_level_density}.
\end{theorem}

\begin{figure}
\centering
\caption{One level density of zeros for even twists with $0<d\leq 400 \:000$ of the $L$-function associated with the elliptic curve $E11.a3$, showing the discrepancy between the zeros data and the ratios conjecture prediction at the origin.}
\label{oldplot_even_zoom}
\includegraphics[width=10cm]{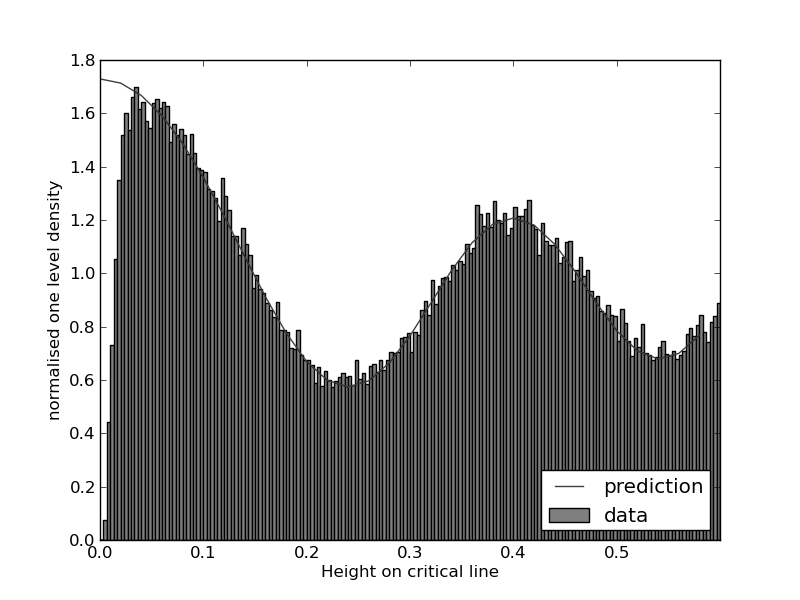}
\end{figure}

Consider the elliptic curve $E11.a3$ (we shall use LMFDB notation \cite{lmfdb} for elliptic curves throughout). The factor multiplying $f(t)$ in the integrand of (\ref{oneleveldensity}) is plotted, as a function of $t$, with $X=400 \:000$, in Figure \ref{oldplot_even_zoom}. The data plotted is a histogram of all zeros up to height 0.6 on the critical line of all $L$-functions in $\mathcal{F}^{+}$ with $0<d\leq400 \:000$.  However, as can be seen in the figure, the one-level density derived from the ratios conjecture does not fit the data well at the origin. The extra zero repulsion from the central point was first noticed by Miller \cite{kn:mil05} in a different family of elliptic curve $L$-functions, and has no analogue in the statistics of $SO(2N)$ eigenvalues. The excised model was conceived in order to model the statistical behaviour of the lowest zero in a family of quadratic twists, capturing Miller's repulsion from the central point. 

 \subsection{The central value formula}
\begin{figure}[h!]
\centering
\caption{The distribution of central values of even twists with $0<d\leq400 \:000$ of the $L$-function associated with $E11.a3$. Peaks from successive integer values of $c_E(\left|d\right|)$ can be clearly seen. The spike at 0 is not plotted as it would distort the scale. }
\label{even1}
\includegraphics[width=10cm]{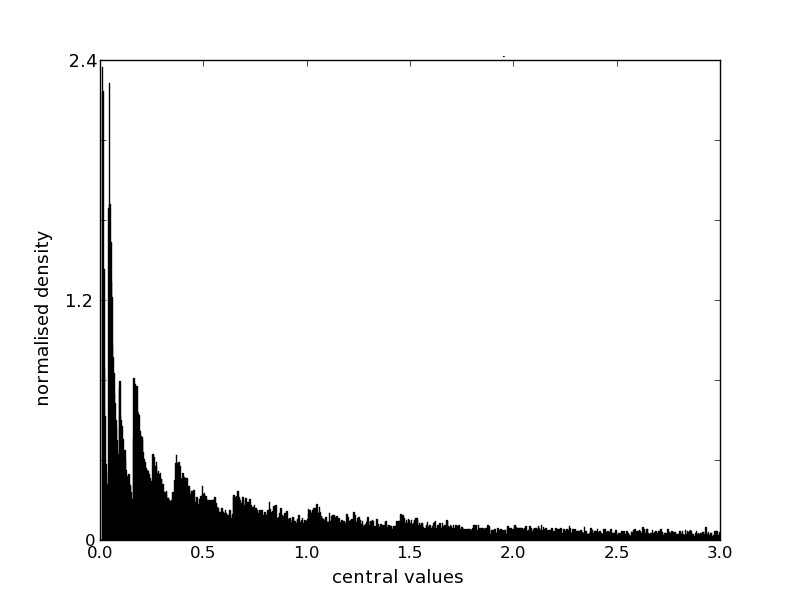}
\end{figure}

\begin{figure}[h!]
\centering
\caption{Close-up of the origin of Figure \ref{even1}, showing the gap betwen 0 and $\kappa_E X^{-1/2}$.}
\label{even2}
\includegraphics[width=10cm]{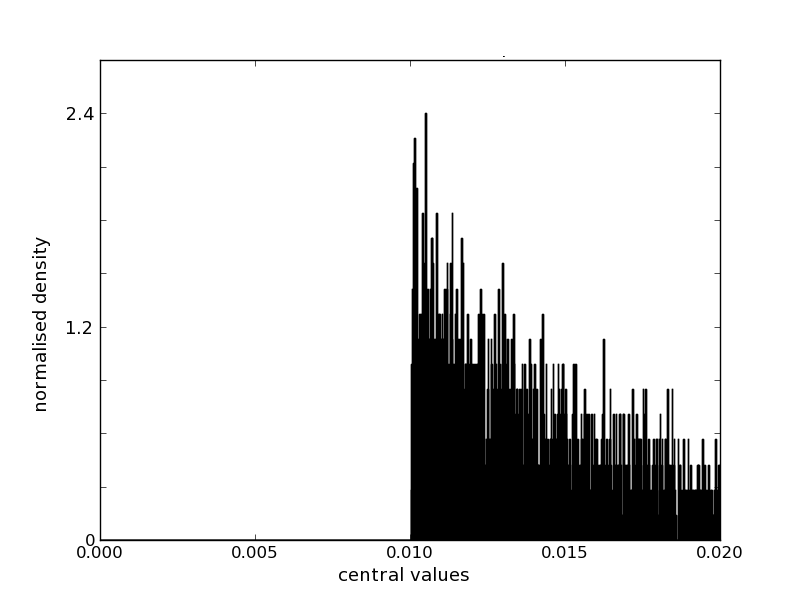}
\end{figure}

The formula of  Kohnen and Zagier \cite{kohnen_zagier} (see also, for example, \cite{kn:waldspurger81,waldspurger,baruch_mao} for related results)  give us a formula for the central value of $L_E(s, \chi_{d})$:
\begin{equation}
\label{critical_value_even}
L_E\left(\frac{1}{2}, \chi_{d}\right) = \kappa_{E} \frac{c_{E} \left( \left| d \right| \right)^{2}}{\sqrt{\left| d \right|} }.
\end{equation}
Here $\kappa_{E}$ is independent of $d$ and so is constant for each family of twists.  (The value of $\kappa_E$ has been computed for many such families by Michael Rubinstein and can be found in Table 3 of \cite{kn:ckrs05}.) Importantly, $c_{E} (\left| d \right|)$ is an integer (specifically, the $c_{E} \left( \left| d \right| \right)$s are the Fourier coefficients of a weight 3/2 modular form). The integer nature of $c_{E}$ means that when we plot the distribution of $ L_E\left(\frac{1}{2}, \chi_{d}\right)$ for a finite range of $0<d \leq X$, we get no values between zero and $\kappa_E X^{-1/2}$. This can be seen in Figures \ref{even1} and \ref{even2}.  Note that in the large $X$ limit the gap between 0 and $\kappa_E X^{-1/2}$ will close.

\begin{figure}[h!]
\centering
\caption{Distribution of central values of even twists with $0< d\leq 400\:000$ of the $L$-function associated with $E11.a3$ compared with the distribution of characteristic polynomial values $\Lambda_B(1)$ for $B\in SO(24)$.}
\label{even3}
\includegraphics[width=10cm]{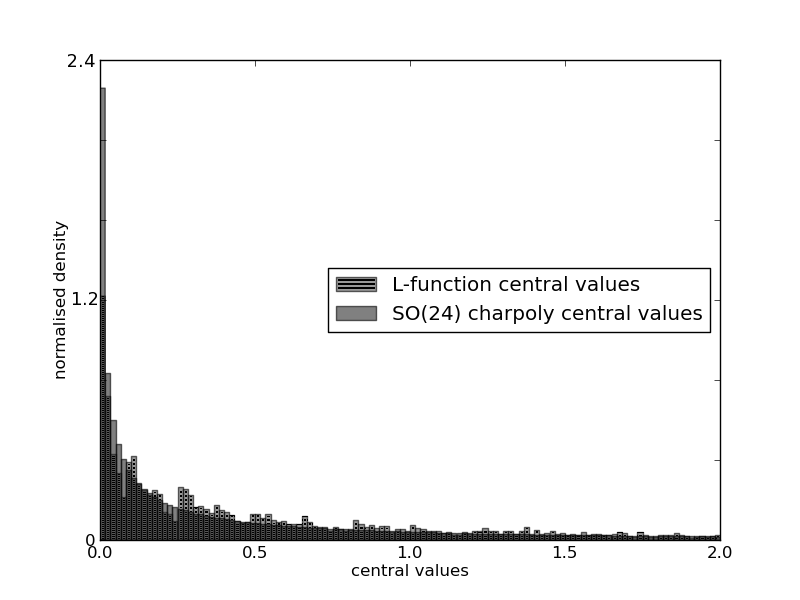}
\end{figure}

As mentioned in the introduction, we expect the distribution of the $L$-function zeros to tend, as $X\rightarrow \infty$,  to the large-$N$ limit of the distribution of eigenvalues from $SO(2N)$.   Previous work \cite{vanishings,kn:dhkms12,one_level_density,kn:huymilmor} leads us to expect that the $L$-function values for this family should be modelled by the values of characteristic polynomials $\Lambda_B(s)$  from $SO(2N)$. The zeros of these characteristic polynomials all lie on the unit circle, which is equivalent to the critical line in the number theory case. The eigenvalues come in complex conjugate pairs, so they have a symmetry around the  point 1 on the unit circle. The point 1 corresponds to the point 1/2 for the family of $L$-functions as the zeros also display  symmetry around this point.  For finite conductor, the distribution of the characteristic polynomial values at the point 1, $\Lambda_B(1)$, appears in some sense to be a smooth approximation to the `choppier' distribution of the  central $L$-function values. This can be seen in Figure \ref{even3}. To determine the appropriate value of $N$ in this figure, we equate the density of zeros near the central point with the average density of eigenvalues: 
\begin{equation}\label{eq:equatingdensities}
\frac{1}{\pi}\log\left(\frac{\sqrt{M}X}{2\pi}\right)=\frac{N}{\pi}.
\end{equation}
Values of $X=400\:000$ and $M=11$ give $N$, to the nearest integer, as 12. 

\subsection{Modelling the statistics of the lowest zero using the excised ensemble}

\begin{figure}[h!]
\centering
\caption{Cumulative distribution of the eigenvalue nearest the point 1 from the $SO(24)$ ensemble and the cumulative distribution of the lowest zero of rank 0 even quadratic twists ($0<d < 400000$, $d$ prime) of the $L$-function associated with $E11.a3$. Here, and in all the other plots of $L$-function zeros, we have scaled each zero by $\log(\frac{\sqrt{M} d}{2 \pi})$ so that each $L$-function has the same mean spacing of zeros near the central point. Similarly, in all figures the random matrix data is scaled so that the means of the two distributions agree.}
\label{even5}
\includegraphics[width=10cm]{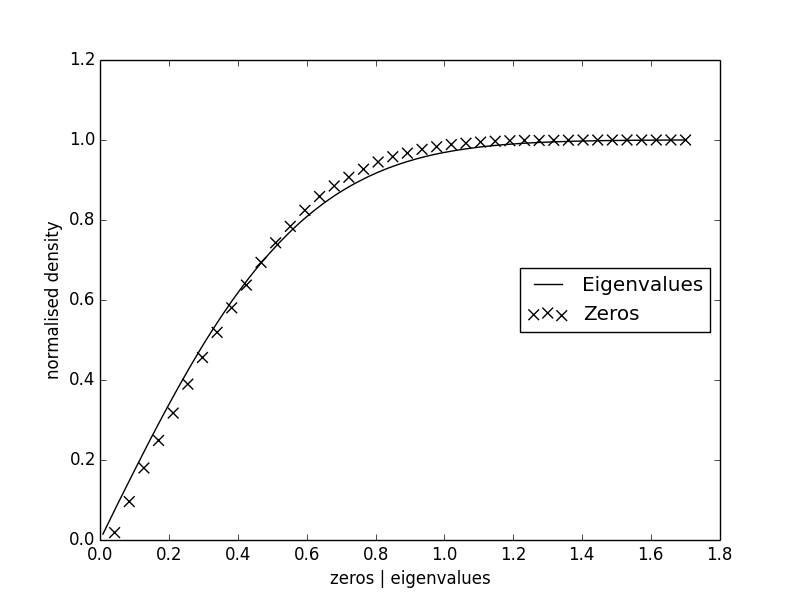}
\end{figure}

The distribution of values of $\Lambda_B(1)$ does not capture the hard gap for finite $X$ shown in Figure \ref{even2}.  The consequence of this can be seen in Figure \ref{even5}: when we look at the distribution of lowest zero (here plotted cumulatively) across the family $\mathcal{F}^{0}$ and compare this with the distribution of the eigenvalue closest to 1 as we sample matrices from the full $SO(2N)$ we see there is not good agreement between the plots. Note that here we are including only $L$-functions in our family that do not have a zero at the point 1/2: those that the Birch and Swinnerton-Dyer Conjecture predicts are associated with rank 0 curves. In addition we have restricted the fundamental discriminants $d$ to prime values as this simplifies the structure of the $c_E(|d|)$s (see \cite{vanishings}). The agreement between the distribution of  first zeros of rank zero twists and the distribution of first eigenvalues will be the measure of the success of the various models.  We will quantify this agreement later in this section.  

\begin{figure}[h!]
\centering
\caption{Cumulative distribution of the eigenvalue nearest the point 1 from the excised $SO(24)$ ensemble with cutoff given at (\ref{eq:RMTcutoff}) and the cumulative distribution of the lowest zero of rank 0 even quadratic twists ($0<d < 400000$, $d$ prime) of the $L$-function associated with $E11.a3$. The fit is significantly better than with the full $SO(24)$ ensemble in Figure \ref{even5}.}
\label{even4}
\includegraphics[width=10cm]{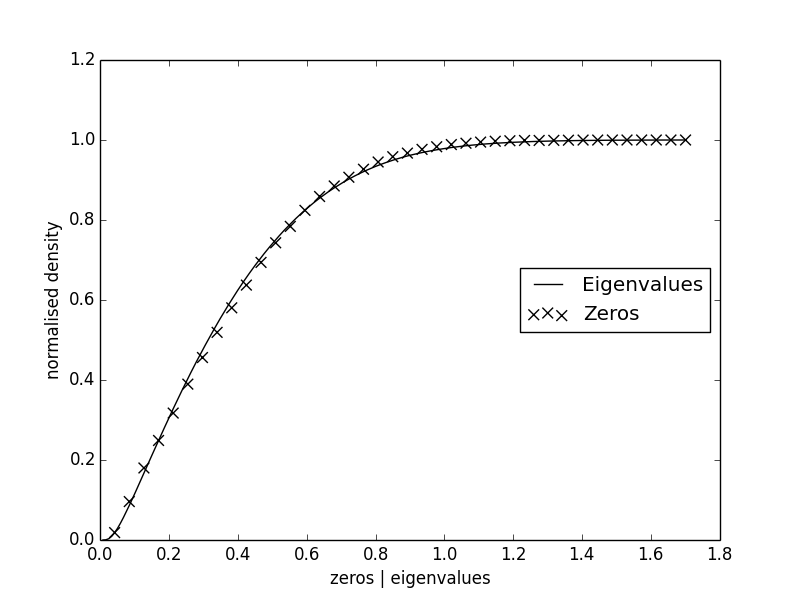}
\end{figure}

We find that if we discard matrices from $SO(2N)$ with characteristic polynomial values $\Lambda_B(1)$ taking a value smaller than a suitably chosen cutoff, then the distribution of the eigenvalue nearest to the point 1 of the resulting excised ensemble is a much improved fit for the distribution of the lowest zero near the point 1/2 of the family $\mathcal{F}^{0}$, which can be seen in Figure 
\ref{even4} for the family of quadratic twists of the elliptic curve $E11.a3$. This was the key observation in \cite{kn:dhkms12}. 

Equation (\ref{critical_value_even}) alone does not quite give us the full picture. An excised subensemble of $SO(2N)$ only gives us a good model for the zero statistics if we adjust the cutoff value to be smaller than the expected $\frac{\kappa_E  } {\sqrt{\left| X \right|} }$. If we write our adjusted cutoff value as $\frac{\kappa_E }  {\sqrt{\left| X \right|} } \delta $, we may calculate $\delta$ by following the method proposed in \cite{kn:dhkms12}. Following the notation of that paper, we define the $s$th moment
\begin{equation}
M_E(X,s) = \frac{1}{X^*} \sum_{\substack{0 < d \leq X \\ d \in \mathcal{D}^+}} (L_E (1/2, \chi_d))^s,
\end{equation}
where $X^* \coloneqq \# \{ 0 < d \leq X  | d \in \mathcal{D}^+\}$. As in \cite{values} we expect, for large $X$ and $N \sim \log X$ (from (\ref{eq:equatingdensities})), that the following moment conjecture holds:
\begin{equation} \label{moments}
M_E(X,s) \sim a_s (E) M_O (N,s),
\end{equation}
where 
\begin{equation}
M_O(N,s) \coloneqq \int _{B \in SO(2N)} \Lambda_B(1)^s \mathrm{d}B  .
\end{equation}
Here, $\mathrm{d}B$ is Haar measure on $SO(2N)$. An expression for the arithmetic factor $a_s(E)$ is given by equation (5.4) in \cite{kn:dhkms12}:
\begin{equation}\label{eq:asEprodLeq}
  \begin{split}
    a_s(E) \ &= \
    \Bigg[\prod_p\Big(1-\frac{1}{p}\Big)^{s(s-1)/2}\Bigg]\\
    &\qquad\times \Bigg[ \prod_{p\nmid M} \frac{p}{p+1} \Bigg(\frac{1}{p}+ \frac{1}{2}
    \Big[ \mathcal{L}_p \Big( \frac{1}{p^{1/2}}\Big)^s +
    \mathcal{L}_p\Big(\frac{-1}{p^{1/2}}\Big)^s\Big]
    \Bigg) \Bigg]\\
    &\qquad\times \mathcal{L}_M\Big(\frac{\pm\omega(E)}{M^{1/2}}\Big)^s,
  \end{split}
\end{equation}
where 
\begin{equation}
{\mathcal L}_p(z)  \ \coloneqq\  \sum_{n=0}^\infty \lambda(p^n)z^n=(1 -
 \lambda(p)z + \psi_M(p)z^2)^{-1}
\end{equation}
and $\psi_M(p)$ is given by
\begin{equation}\label{psi}
\psi_{M}(p) =
\begin{cases}
1 \text{~if~} p \nmid M\\
0 \text{~otherwise}.
\end{cases}
\end{equation}
Equation \eqref{eq:asEprodLeq} holds for prime conductor $M$, where $\omega(E)$ is the sign of the functional equation of~$L_E(s)$.  The $\pm$ in the last line of \eqref{eq:asEprodLeq} is $+$, if the family involves  twists by positive fundamental discriminants $0<d\leq X$. The $-$ sign corresponds to twists by negative fundamental discriminants $-X\leq d<0$.  

From \cite{values}, we have 
\begin{equation}
M_O(N,s) = 2^{2Ns} \prod _{j = 1} ^{N} \frac{\Gamma(N + j - 1) \Gamma(s + j - 1/2)}{\Gamma(j - 1/2) \Gamma(s + j + N - 1)} .
\end{equation}
We want to compare the value distributions of $\Lambda_B(1)$ and $L_E(1/2,\chi_d)$.  These can be found from the moments by using the Mellin transform. For example, if we consider a variable $y$ representing the possible values taken by $\Lambda_B(1)$, the probability density, $P_O(N,y)$, for $y$ is given by:
\begin{align}\label{mellin}
P_O(N,y)\mathrm{d}y \ \coloneqq & \ \mathrm{Prob}(y \leq \Lambda_B(1) < y + dy | B \in SO(2N)) \nonumber \\
= & \ \frac{1}{2 \pi \mathrm{i} y } \int _{c - \mathrm{i} \infty} ^{c + \mathrm{i} \infty} M_O(N,s) y ^{-s} \mathrm{d}s \ \mathrm{d}y .
\end{align}
For small $y$, \cite{vanishings} and then \cite{kn:dhkms12} show that the  main contribution to the integral in (\ref{mellin}) comes from the simple pole at $s = -1/2$, giving
\begin{equation}\label{eq:PO}
P_O(N,y)\mathrm{d}y \sim y^{-1/2} h(N) \mathrm{d}y,
\end{equation}
where 
\begin{align}
h(N)   \coloneqq &  \underset{s = -1/2}{\operatorname{Res}} M_O(N,s) \nonumber \\
	=& \  2^{-N} \Gamma(N)^{-1} \prod _{j = 1}^N \frac{\Gamma(N + j - 1)\Gamma(j)}{\Gamma(j - 1/2) \Gamma(j + N - 3/2)} .
\end{align}
For large $N$, using the definition of the Barnes $G$-function in \cite{G-function}, we have the asymptotic 
\begin{equation}\label{gammaasymptotic}
h(N) \sim 2^{-7/8} G(1/2) \pi^{-1/4} N^{3/8}.
\end{equation}
From this we can find, for small $\rho$, an asymptotic expression for 
\begin{align}
\mathrm{Prob} (0 \leq \Lambda_B (1) \leq \rho) &= \int _{0} ^{\rho} P_O (N,y) \mathrm{d}y \nonumber \\
&\sim \int _{0} ^{\rho} y^{-1/2} h(N) \mathrm{d}y \nonumber \\
& = 2 \rho^{1/2} h(N) .
\end{align}
Now due to (\ref{moments}), we expect that there exists a smooth approximation to the probability density for the $L$-function values for our familiy, which we shall call $P_E(d,\tilde{y})$. This distribution should satisfy
\begin{align}\label{eq:PEPO}
P_E(d,\tilde{y}) \coloneqq &\frac{1}{2 \pi \mathrm{i} \tilde{y}} \int_{c - \mathrm{i} \infty} ^{c + \mathrm{i} \infty} a_s(E) M_O (\log d, \tilde{y}) \tilde{y}^{-s}  \mathrm{d}s\nonumber \\
\sim &  \ a_{-1/2}(E)  P_O(\log d, \tilde{y})
\end{align}
for small $\tilde{y}$. The central value formula (\ref{critical_value_even}) implies that
\begin{equation}
\left( L_E(1/2,\chi_d) < \frac{\kappa_E}{\sqrt{d}} \right) \implies \left( L_E(1/2,\chi_d) = 0 \right).
 \end{equation}
Thus we might expect that for a random variable $\tilde{y}$ distributed with probability density $P_E(d,\tilde{y})$,
\begin{equation} \label{wouldbenice}
\displaystyle \sum_{\substack{d \in \mathcal{D}^+ \\0 <  d < X}} \mathrm{Prob} \left(\tilde{y} < \frac{\kappa_E}{\sqrt{d}} \right) \substack{?\\=} \frac{\#\{L_E(1/2,\chi_d) = 0 \ | \ d < X, d \in \mathcal{D}^+ \}}{\#\{d<X \ | \ d \in \mathcal{D}^+\} },
\end{equation}
where the right hand side, according to the Birch and Swinnerton-Dyer Conjecture, is the proportion of higher order curves in our family, obtained by simply counting how many curves have $L_E(1/2,\chi_d) = 0$.

The model (\ref{wouldbenice}) was proposed in \cite{vanishings} and resulted in a prediction for the number of $L$-functions taking the value 0 at the central point in the family of even quadratics twists that appears, from intensive numerical testing, to have the correct asymptotic dependence on $X$ but with the wrong overall constant.  

This is the reason that in \cite{kn:dhkms12} it was proposed that instead of taking $\tilde{y}<\frac{\kappa_E}{\sqrt{d}}$, one should instead define an `effective' cutoff, $\delta \frac{ \kappa_E}{\sqrt{d}}$.  The value of $\delta$ is found by calculating an asymptotic expression for the number of $L$-functions in our family that are zero at the point 1/2, modelled by  $P_E(d,\tilde{y})$ as in the the left hand side of (\ref{wouldbenice}), but with $\tilde{y}<\frac{\delta \kappa_E}{\sqrt{d}}$, and identifying this with Michael Rubinstein's numerical determination of the same quantity.  From here on we consider just $d\in \mathcal{D}^+$ that are prime since these numerical results were calculated for prime fundamental discriminants only.  That is, from the left side of (\ref{wouldbenice}) (but with $\tilde{y}<\frac{\delta \kappa_E}{\sqrt{d}}$) we have, following \cite{kn:dhkms12}:
\begin{align}\label{asymptotic_equation}
&\#\{L_E(1/2,\chi_d) = 0\ |d \ \mathrm{prime}, \ d < X, d \in \mathcal{D}^+   \} \nonumber \\
&\sim \frac{1}{4 \log X} 2 a_{-1/2}(E) \sqrt{\kappa_E} 2^{-7/8} G(1/2) \pi^{-1/4} (\log X)^{3/8} \delta^{1/2} \frac{4}{3} X^{3/4}.
\end{align}
If we divide both sides of (\ref{asymptotic_equation}) by $a_{-1/2} (E) \sqrt{\kappa_E} X^{3/4} (\log X)^{-5/8}/4$, numerical investigations by Michael Rubinstein counting the number of $L$-functions that have vanishing central value  have found that the left hand side approaches a constant as we increase $X$. For $E11.a3$, this constant is approximately 0.2834620. This gives 
\begin{align}\label{computedelta}
\frac{8}{3} 2^{-7/8} G(1/2) \pi^{-1/4} \delta^{1/2} &\approx  0.2834620 \nonumber \\
\implies  \delta & \approx 0.185116.
\end{align}

Thus the correct version of (\ref{wouldbenice}) for the family of quadratic twists of $E11.a3$ with positive $d$ is
\begin{equation} \label{eq:isnice}
\displaystyle \sum_{\substack{d \in \mathcal{D}^+ \\0 <  d < X}} \mathrm{Prob} \left(\tilde{y} < \frac{\delta\kappa_E}{\sqrt{d}} \right) = \frac{\#\{L_E(1/2,\chi_d) = 0 \ | \ d < X, d \in \mathcal{D}^+ \}}{\#\{d<X \ | \ d \in \mathcal{D}^+\} },
\end{equation}
where $\delta$ is given by (\ref{computedelta}).

In \cite{kn:dhkms12} the authors use the cutoff $\delta \kappa_E X^{-1/2}$ to produce a working model for the lowest zero of rank zero $L$-functions in a family of even quadratic twists.  Since $P_E(d,\tilde{y})$ is conjectured to be just $a_{-1/2}(E)$ times the distribution of characteristic polynomials $P_O(N,\tilde{y})$ for very small values of $\tilde{y}$ (see (\ref{eq:PEPO})), and the distributions vary like $\tilde{y}^{-1/2}$ (see (\ref{eq:PO})), the constant in the cutoff appropriate for $P_E(d,\tilde{y})$, $\delta \kappa_E$, is scaled to 
\begin{equation}\label{eq:RMTcutoff}
c=a_{-1/2}^{-2}(E)\delta \kappa_E
\end{equation}
 when applied to $P_O(N,\tilde{y})$.  Since (\ref{eq:isnice}) implies that matrices with $\Lambda_B(1)$ taking a value below this cutoff are involved with modelling the number of $L$-functions that are zero at the central point, and therefore by Birch and Swinnerton-Dyer are presumed to be associated with an elliptic curve having rank 2 or greater, the matrices with $\Lambda_B(1)$ greater than the cutoff can be used to model the $L$-functions that don't vanish at the central point: those associated with a curve of rank 0.   Thus the excised model applied in \cite{kn:dhkms12} involves retaining only those matrices from $SO(2N)$ that have $\Lambda_B(1)\geq c \exp(-N/2)$. $N$ is given by (\ref{eq:equatingdensities}). 

\begin{figure}[htbp]
\begin{center}
\includegraphics[scale=.28,
angle=0]{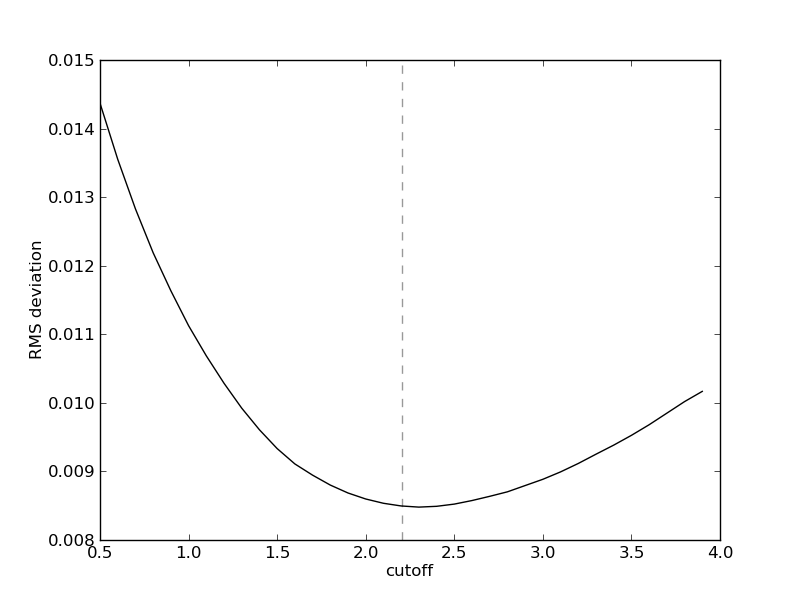}\hspace{0.05in}
\includegraphics[scale=.28,
angle=0]{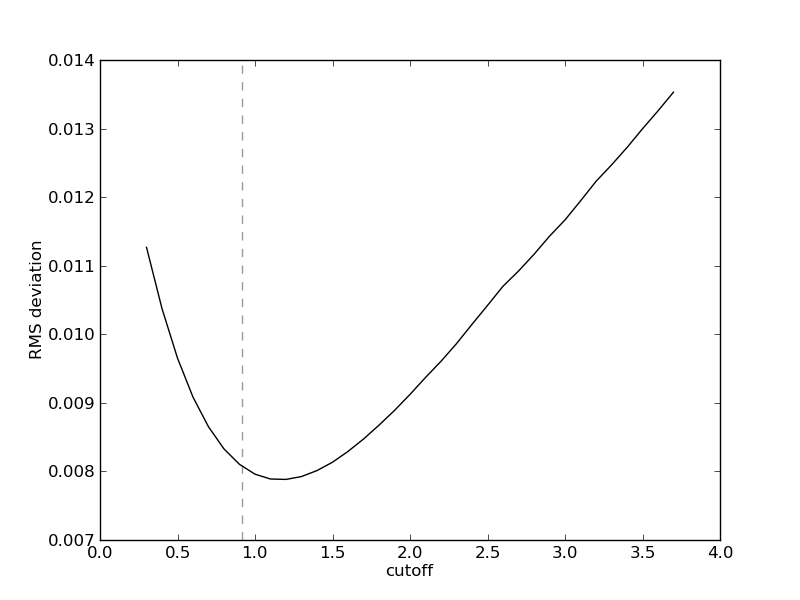}\\
\includegraphics[scale=.28,
angle=0]{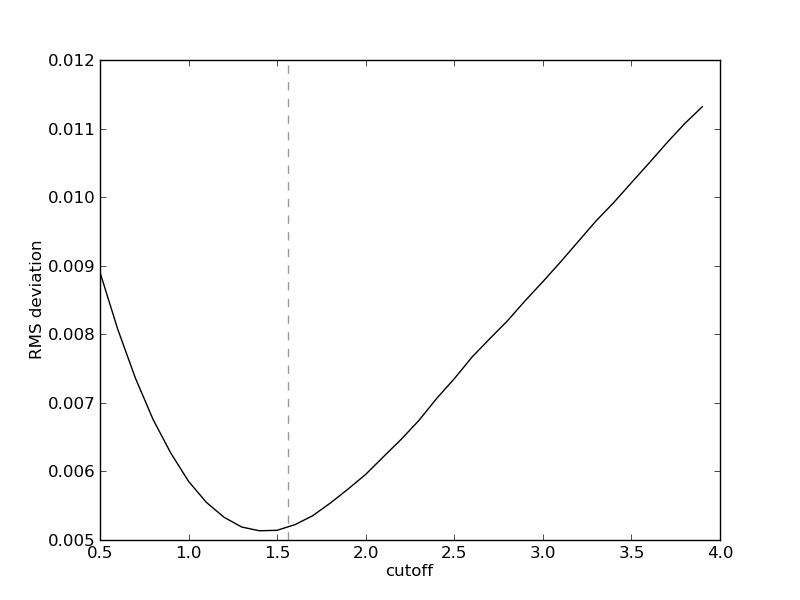}\hspace{0.05in}
\includegraphics[scale=.28,angle=0]{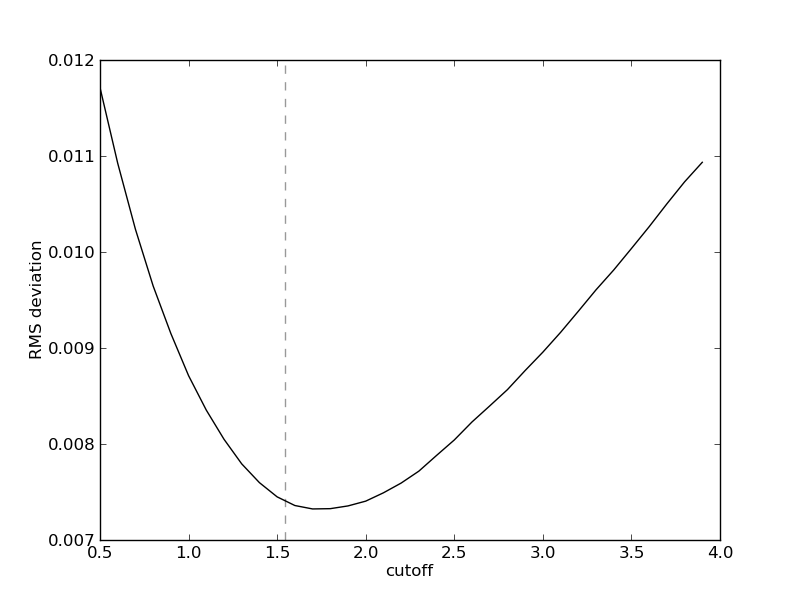}
\caption{RMS deviation between the distribution for the first zero of rank 0 curves in a family of quadratic twists and the distribution of the eigenvalue nearest 1 for the excised model as the cutoff varies.  Here we have used $X = 308317$ when $M = 11$ and $X = 234605$ when $M = 19$ so that the equivalent matrix size is as close to 12 as possible.   The value of the cutoff given by (\ref{eq:RMTcutoff}) in each case is plotted as a vertical line. For a sense of scale note that a cutoff equal to zero is the full $SO(2N)$ ensemble. Top left: twisting $E11.a3$ with positive fundamental discriminants. Top right: negative twists of $E11.a3$.  Bottom left: positive twists of $E19.a2$.  Bottom right: negative twists of $E19.a2$. } \label{fig:RMSdeviations}
\end{center}
\end{figure}

It is worth noting here what effect this has when $N$, as defined in (\ref{eq:equatingdensities}), does not turn out to be an integer.  Although in evaluating $c \exp(-N/2)$ we can use a non-integer value of $N$,  we are forced to use integer size matrices when we generate matrices from $SO(2N)$ to test the model. To remove the effect of computing with a matrix size that  is slightly different from the value of $N$ given by (\ref{eq:equatingdensities}), for the plots in Figure \ref{fig:RMSdeviations} we have chosen $X$ differently for each family such that in each case $N$ is as close to an integer as possible.

 The effectiveness of the excised model with cutoff (\ref{eq:RMTcutoff}) in predicting the distribution of the lowest zero for $L$-functions associated with rank 0 curves can be seen in Figure \ref{even4}.  Note that as in \cite{kn:dhkms12} before comparing the distribution of first eigenvalues with the distribution of lowest zeros, the two distributions are scaled to have the same mean.  This is to account for the fact that the random matrix model contains no number theoretical information, so although it appears to predict correctly the shape of the distribution, there is a slight difference in the means. The improvement of the excised model in Figure \ref{even4} over the full $SO(2N)$ ensemble shown in Figure \ref{even5} is clear. This can be quantified by considering the root-mean-square (RMS) deviation between the bin heights for the excised model and the bin heights for the number theoretic data in the histograms in Figure \ref{even4}. The better our model, the smaller this quantity will be. It was checked in \cite{kn:dhkms12} for twists (with positive $d$) of the curve $E11.a3$, and in later investigations for three other families of even quadratic twists, that as we vary the cutoff, the best match (i.e. the smallest RMS deviation) is for a cutoff very close to the value $c$. Figure \ref{fig:RMSdeviations} plots the RMS deviation for each of these families as the value of the cutoff varies.  The vertical line indicates the position of the value $c$ determined by (\ref{eq:RMTcutoff}) for each family. 
 
Thus, the excised ensemble provides a simple model for the distribution of the first zero of even, prime quadratic twists of an elliptic curve $L$-function that captures Miller's repulsion from the central point.  It also provides a means of predicting some significant information about the number of higher rank curves in such a family in the sense that it results in a conjecture that the number of these curves is proportional to $X^{3/4}(\log X)^{-5/8}$.  However it leaves us with the mysterious fact that the most effective cutoff is $\delta\kappa_E/\sqrt{d}$ rather than $\kappa_E/\sqrt{d}$.  In the next section we adapt the excised model to shed light on the role of $\delta$. 

\section{Beyond the excised model}

\subsection{The inverse cubic model}\label{sect:modweightmod}

Despite the success of the excised model, there are some problems with its interpretation. For modelling $L$-function zeros in the regime where the conductor is not large, it seems strange that the excised model cutoff (that gives the correct number of vanishing $L$-functions) is not at the most obvious value of $\kappa_E X^{-1/2}$, but is instead at a point significantly closer to zero. If the distribution of $SO(2N)$ characteristic polynomial values were a very good smooth approximation to the distribution of $L$-function values, then it would seem counterintuitive that (\ref{wouldbenice}) would not hold. 

\begin{figure}[h!]
\centering
\caption{Distribution of the central values of $L$-functions associated with even quadratic twists of $E11.a3$ ($0<d < 400000$, $d$ prime). This figure includes only those $L$-functions for which $c_E(\left|d\right|) = 1$. This is the left-most peak visible in Figure \ref{even1}. A best fit line proportional to $\frac{1}{\tilde{y}^3}$ is also plotted for comparison.}
\label{first_spike}
\includegraphics[width=10cm]{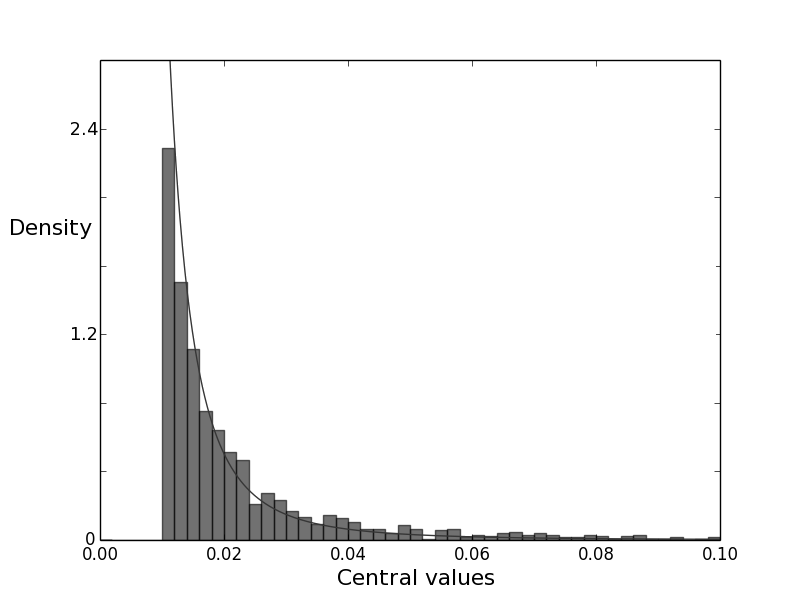}
\end{figure}

If we look at Figure \ref{even3}, we can see that there is some structure in the number theory data that the excised model does not take into account. The $L$-function values are given by (\ref{critical_value_even}) and their density histogram has a `peak' for each integer value taken by $c_{E} \left( \left| d \right| \right)$. This peak is not a delta spike, but is  `smeared out' by the factor $\frac{1}{\sqrt{\left|d\right|}}$ in (\ref{critical_value_even})  as $d$ runs over fundamental discriminants between $0$ and $X$. Figure \ref{first_spike} shows the peak due only to the $L$-values for which $c_{E} \left( \left| d \right| \right) = 1$, illustrating the sharp left edge corresponding to the largest value of $d=X$ as well as the decay to the right corresponding to smaller values of $d$. 

Since the excised model (which only takes into account the hard gap at the origin and not this series of peaks) is so much more successful than the full $SO(2N)$ ensemble, it is natural to ask if a modification to the excised model which does include information about these peaks would improve our model for the zero statistics still further. 

If we think of excision as imposing a Heaviside step function weight on $SO(2N)$, the logical extension of this method is to impose a more complicated weight on $SO(2N)$ in such a way as to select a subensemble with a distribution of values of $\Lambda_B(1)$ which closely matches the distribution of values of $L_E(1/2,\chi_d)$. (Note that as we are now modelling details of the peaked distribution of $L$-function values, we are not using the approximation (23) that holds for small $L$-values and involves scaling $P_O(N,y)$ by the arithmetic factor in order to obtain a smooth approximation to the $L$-value distribution, as described at (29).) Selecting a subensemble of $SO(2N)$ with a chosen distribution of values $\Lambda_B(1)$ may be achieved computationally by generating Haar-distributed orthogonal matrices following Mezzadri \cite{kn:mezzadri07}, and then using von Neumann's selection-rejection method on the resulting data. 

This selection-rejection algorithm goes through the computer generated set of $SO(2N)$ matrices one at a time and selects matrices for our modified ensemble with a probability  determined by the value at 1 of the characteristic polynomial.  We shall refer to this probability as $W(N,y)$, where $y$ is again a variable representing the  possible values of the characteristic polynomials. Thus if a randomly generated matrix  $B\in SO(2N)$ has a value at 1 of $\Lambda_B(1)$, that matrix is retained with a probability $W(N,\Lambda_B(1))$ and is otherwise discarded. This gives us a random matrix ensemble whose central values are distributed like $P_{\mathrm{ic}}(N,y) \coloneqq N_W W(N,y)P_O(N,y)$, where as before $P_O(N,y)$ is the probability distribution of central values from the full $SO(2N)$ ensemble, and $N_W$ is a normalisation constant such that
\begin{equation} \label{eq:N_W}
N_W\int _{0}^\infty W(N,y) P_O(N,y)\mathrm{d}y = 1,
\end{equation}
so that $P_\mathrm{ic}(N,y)$ is a proper probability distribution.

We note that although we call it a `probability', it is not necessary to require that $W(N,y) < 1$. If $W(N,\Lambda_B(1)) >\ $1 for a random $SO(2N)$ matrix $B$, we can implement this by adding $\lfloor W(N,\Lambda_B(1)) \rfloor$ copies of that matrix to our modified ensemble, and then adding a further copy with probability $\{ W(N,\Lambda_B(1)) \}$ (where $\lfloor \rfloor$ denotes the integer part and $\{ \ \}$ the fractional part).

As explained below, we shall be interested in the case where $W(N,y)$ is flat over most of its range, so we shall set $W(N,y)=1$ over the range in which it is flat. Thus the average, $\langle W(N,y) \rangle$ of $W$ over the full range of $y$ will be $\langle W(N,y) \rangle \approx 1$. This is  computationally efficient  because if $\langle W(N,y) \rangle < 1$, then as we increase $\langle W(N,y) \rangle$, we reject fewer $SO(2N)$ matrices from our ensemble than we need to and thus waste less computer time. If $\langle W(N,y) \rangle > 1$, then we do not gain any more accuracy by increasing $\langle W(N,y) \rangle$ since we do not actually gain any accuracy in our data sets from adding multiple copies of every matrix to our ensemble. Therefore it makes sense to have $\langle W(N,y) \rangle $ approximately equal to 1. Since $W(N,y)$ will not be identically 1, then for some range of $y$, $W(N,y)$ will be greater than one, as mentioned above.

\subsubsection{Finding $W(N,y)$}\label{sect:modifiedweightmodel}

\begin{figure}[h!]
\centering
\caption{An example of a weight $W(N,y)$ which we might wish to be imposed on top of Haar measure. The left-most peak would start at $y=\kappa_E  {X}^{-1/2}$ and the second discontinuity occurs at $ 4\kappa_{E} X^{-1/2}$, where $X$ is the largest value of $d$ in the set of $L$-function data. The shape of the first peak is fixed, but its height depends on the parameter $A$ introduced in (\ref{eq:P_ic_definition}).}
\label{weighty}
\includegraphics[width=10cm]{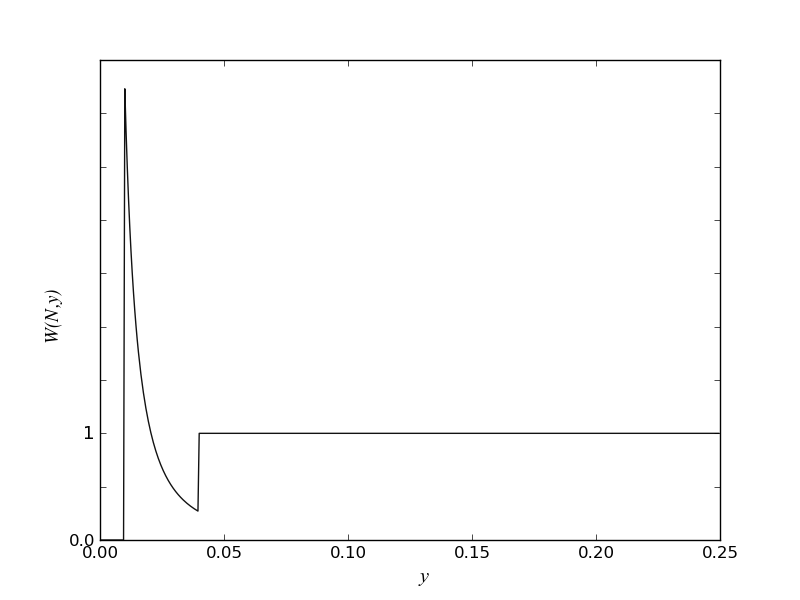}
\end{figure}

The goal here is to find the weight function $W(N,y)$ that we wish to impose upon our Haar-distributed matrix ensemble $SO(2N)$. We will still exclude all matrices whose characteristic polynomial value is less than a cutoff, but in contrast to the excised model, we shall use the `natural' cutoff at $\kappa_E X^{-1/2}$. The start of the second peak should therefore be at $4\kappa_E X^{-1/2}$ (since $L_E(1/2,\chi_d) =\kappa_E c_E(\left|d\right|)^2/\sqrt{d}$ and the first two peaks correspond to $c_E(\left|d\right|)^2 = 1$ and $c_E(\left|d\right|)^2 = 4$ respectively). Since we are most interested in the structure of the distribution for small values (where the discretisation in (\ref{critical_value_even}) is most noticable), we shall choose $W(N,y)$ such that the distribution of characteristic polynomials is similar to the $L$-function value distribution over the entirety of the first peak, and then simplifies to Haar measure for $\Lambda_B(1) > 4 \kappa_E X^{-1/2}$. A sketch of $W$ can be seen in Figure \ref{weighty}. The shape of the peak is found numerically by requiring that when the weight $W(N,y)$ is imposed upon Haar measure, the resulting distribution of $\Lambda_B(1)$ should be an inverse cubic curve (motivated by the fit in Figure \ref{first_spike} and justified later in this section). This defines $W(N,y)$ up to a constant, which shall be discussed in Section \ref{sect:A}.

\begin{figure}[h!]
\centering
\caption{Distribution of prime fundamental discriminants from $(0 < d < 308317)$ for which $c_E(|d|)=1$. The right hand side of the distribution is approximately flat, which helps to justify our assumptions in the calculation of the shape of the first peak.}
\label{dees}
\includegraphics[width=10cm]{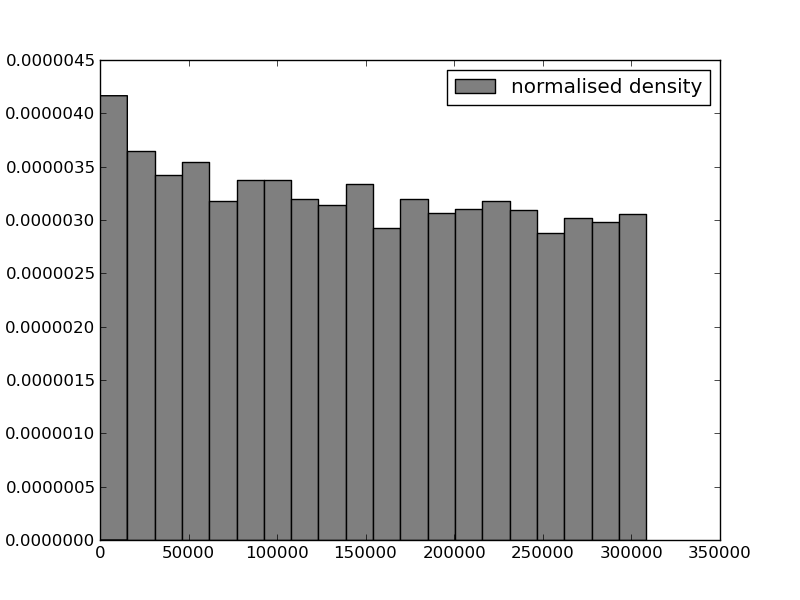}
\end{figure}

We wish to find the functional form of the first peak in the distribution of $L$-values:  see Figure \ref{first_spike}. As before, $\tilde{y}$  represents the values taken by $L(1/2,\chi_d)$. In order to do this, we use the central value formula (\ref{critical_value_even}). We note that for $L$-functions that contribute to the first peak, $L_E(1/2,\chi_d) = \kappa_E d^{-1/2}$. Consider the number of curves with central values in the interval between $\tilde{y}$ and $\tilde{y} + \Delta \tilde{y}$. We have 
\begin{equation}
\tilde{y} \sqrt{d} \leq \kappa_E \leq (\tilde{y} + \Delta \tilde{y}) \sqrt{d}
\end{equation}
so
\begin{equation}
\frac{\kappa_E^2}{(\tilde{y}+ \Delta \tilde{y})^2 }\leq d \leq \frac{\kappa_E^2}{\tilde{y}^2}.
\end{equation}
  For simplicity we make the assumption that the values of $d$ corresponding to $c_E(\left|d\right|) = 1$  are uniformly distributed among the integers in the interval $[0,X]$ (although the distribution of the values taken by $c_E(\left|d\right|)$ is a complicated subject, which may be treated in future work by Michael Rubinstein). As can be seen in Figure \ref{dees}, this is not the case, but the density of these $d$s is slowly varying for most of the range of $d$ that we are considering, particularly for larger $d$.  These larger values of $d$ determine the initial sharp fall of the peak in Figure \ref{first_spike}, so this seems a reasonable approximation to make in order to obtain an approximate shape for the first peak.  Accepting this assumption implies that the number of values of $d$ in an interval is proportional to the length of the interval. As such, the number of curves with central values in the interval $[\tilde{y},\tilde{y} + \Delta \tilde{y}]$ is proportional to 
\begin{equation}
\frac{   \kappa_E^2 } { \tilde{y}^2}    -   \frac{   \kappa_E^2 } { (\tilde{y} + \Delta \tilde{y})^2}  = \Delta \tilde{y} \frac{  \kappa_E^2 ( 2\tilde{y} + \Delta \tilde{y}) } { \tilde{y}^2(\tilde{y}+\Delta \tilde{y})^2}
\end{equation}
if $\tilde{y} > \kappa_E X^{-1/2}$, and 0 otherwise. For $\Delta \tilde{y} \ll \tilde{y}$, this gives us 
\begin{equation}
 \#\{d : \tilde{y} \leq L_E(1/2,\chi_d)  \leq \tilde{y}+\Delta \tilde{y}, c_E(|d|)^2=1\} \sim \frac{2 \kappa_E^2 \Delta \tilde{y}}  {\tilde{y}^3} \nonumber
\end{equation}
so we conclude that a smooth approximating distribution for the first peak in the $L$-value distribution should be an inverse cubic curve. 

Note that the shape of $W(N,y)$ between $\kappa_E X^{-1/2}$ and $4\kappa_E X^{-1/2}$ is not itself an inverse cubic because $W(N,y)$ is the weight that is applied to Haar-distributed matrices in order to ensure a value distribution for $\Lambda_B(1)$ that has inverse cubic shape. As Haar measure does not result in a flat distribution of $\Lambda_B(1)$ in this region, $W(N,y)$ has a more complicated shape. 

Thus we have an argument that the first peak of the distribution of central $L$-values should have a sharp turn on at $\kappa_EX^{-1/2}$ followed by an inverse cubic decay. Our proposed random matrix model is a subensemble of $SO(2N)$ chosen such that the distribution of characteristic polynomials $\Lambda_B(1)$ obeys just such a distribution law between $0$ and $4\kappa_E X^{-1/2}$, and matrices with characteristic polynomials greater than $4\kappa_E X^{-1/2}$ are chosen with Haar measure. This will give us a probability distribution for the central values defined by:
\begin{equation} \label{eq:P_ic_definition}
P_{\mathrm{ic}}(N,y) = \left\{
\begin{alignedat}{2}
&0  &&\text{~if~} 0 <y < \kappa_E X^{-1/2} \\
&N_W Ay^{-3} &&\text{~if~} \kappa_E X^{-1/2}  \leq y < 4 \kappa_E X^{-1/2} \\
&N_W P_O(N,y) && \text{~otherwise.}
\end{alignedat} \right.
\end{equation}
From this, we find
\begin{equation} \label{P_ic}
W(N,y) = \left\{
\begin{alignedat}{2}
&0 &&\text{~if~} 0 < \kappa_E X^{-1/2} \\
&A y^{-3} / P_O(N,y) &&\text{~if~} \kappa_E X^{-1/2}  \leq y < 4 \kappa_E X^{-1/2} \\
&1 &&\text{~otherwise.}
\end{alignedat} \right.
\end{equation}
For this to be useful, we need to be able to calculate values for $P_O(N,y)$. This can be done numerically by generating a large number of $SO(2N)$ matrices, plotting a histogram of the value distribution, and then fitting a high degree polynomial to the resulting set of bin heights.

\subsubsection{The effectiveness of the inverse cubic model}\label{sect:effectiveness}

\begin{figure}[h!]
\centering
\caption{Cumulative plot of the distribution of the first zero of even quadratic twists ($0<d < 308317$, $d$ prime) of $L$-function associated with $E11.a3$, with the distribution of the first eigenvalue from the inverse cubic model using parameter value $A_1$ (give at (\ref{eq:A1}). By eye it is difficult to distinguish between this and the equivalent plot for the excised model (Figure \ref{even4}), but the RMS deviation between these two curves is around 5.8\% lower.}
\label{mod_weight_zeros_eigs}
\includegraphics[width=10cm]{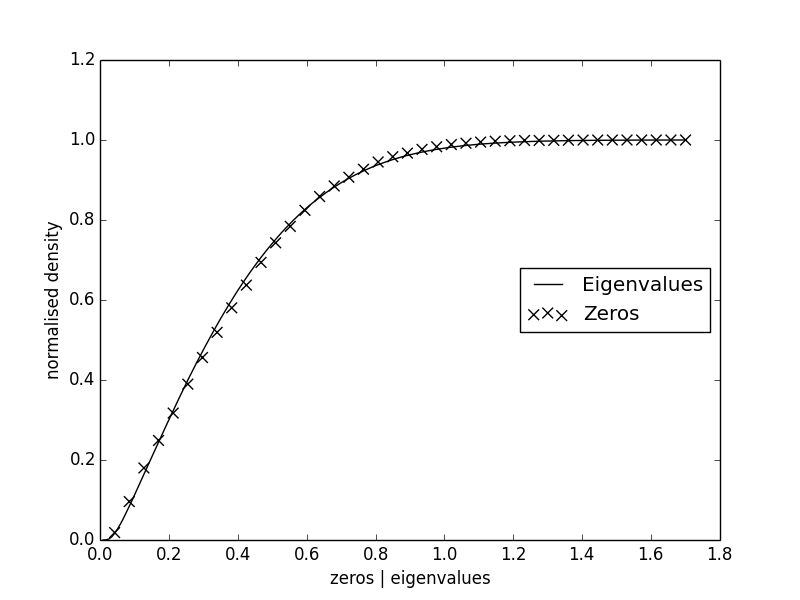}
\end{figure}

\begin{figure}[h!]
\centering
\caption{Finding the optimum height of the first peak for the inverse cubic model, with horizontal line showing the RMS deviation of the excised model. The values of $A_1$ as calculated in Section \ref{sect:A} are marked with vertical lines. Top left: twisting $E11.a3$ with positive fundamental discriminants; showing a 5.8\% improvement on the excised model. Top right: negative twists of $E11.a3$; 6.8\% improvement.  Bottom left: positive twists of $E19.a2$; 5.7\% improvement.  Bottom right: negative twists of $E19.a2$; 12.1\% improvement.}
\label{bsearch}
\includegraphics[width = 6cm]{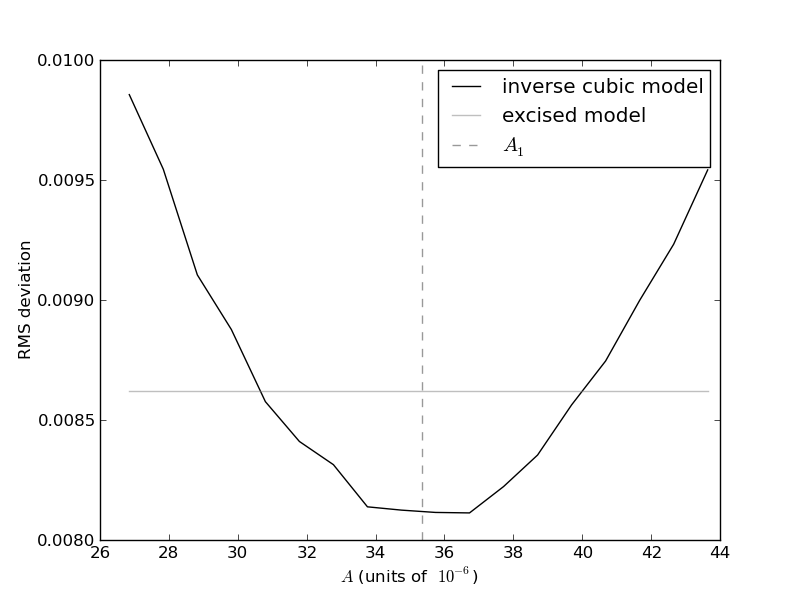}
\includegraphics[width = 6cm]{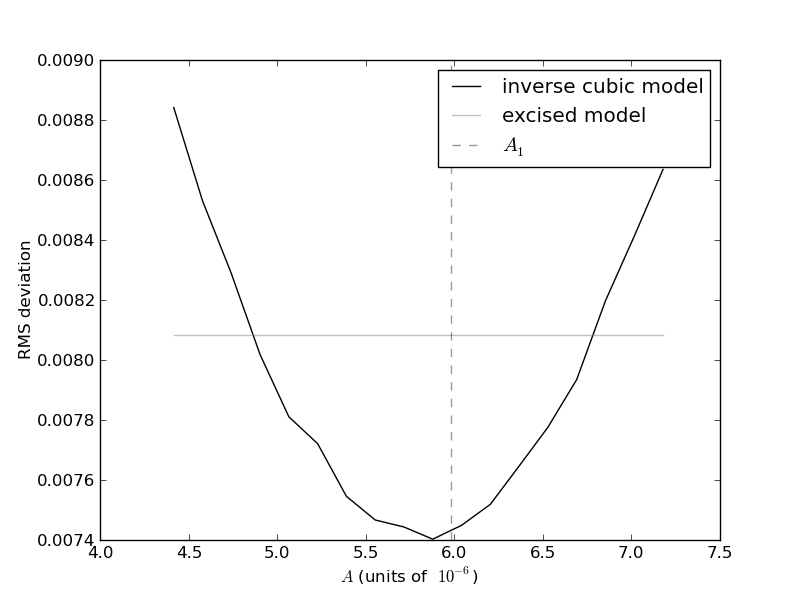}
\includegraphics[width = 6cm]{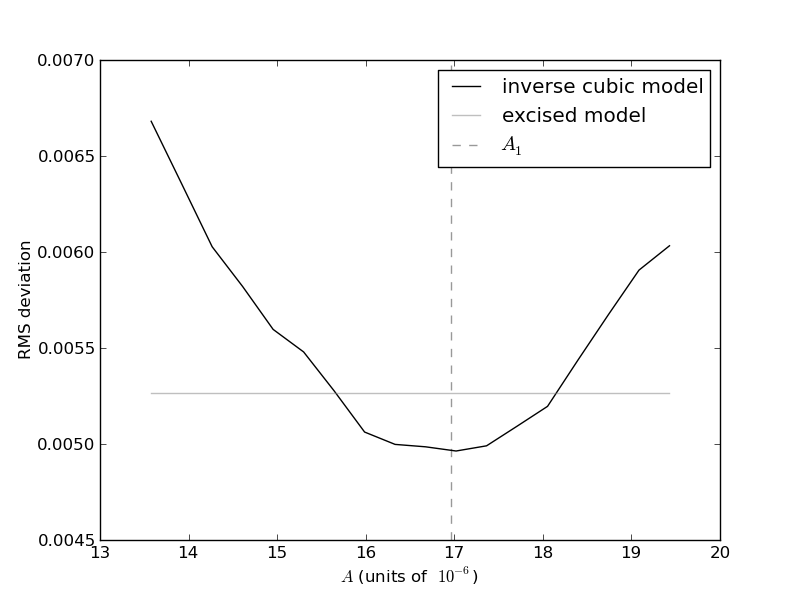}
\includegraphics[width = 6cm]{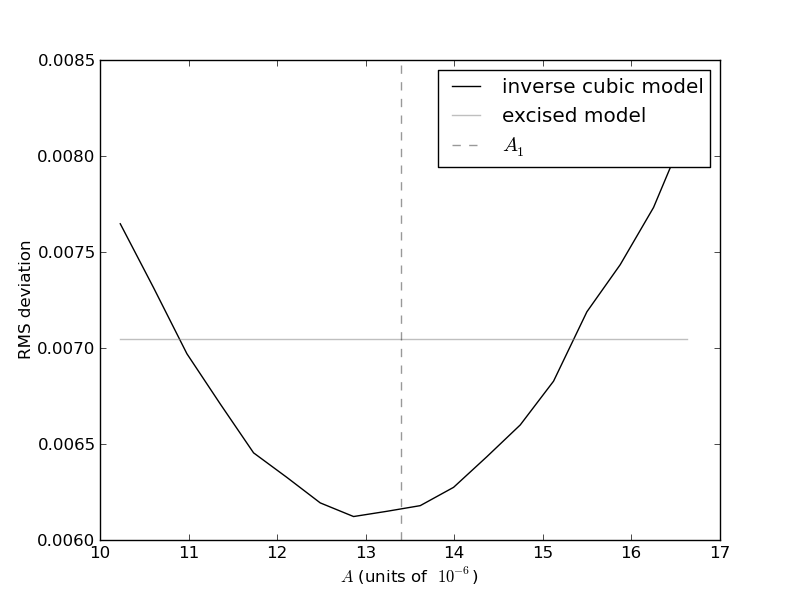}
\end{figure}

As can be seen in Figure \ref{mod_weight_zeros_eigs}, we find that the inverse cubic model described in Section \ref{sect:modifiedweightmodel} does indeed give us a matrix ensemble whose first eigenvalues are a good model for the lowest zeros of rank 0 quadratic twists. In this figure the parameter $A$ is set to $A_1$, as defined at (\ref{eq:A1}).  The predicted value $A_1$ is very close to the value of $A$ that gives the best fit between the two curves in Figure \ref{mod_weight_zeros_eigs}.  This optimal value of $A$ can be found by considering the RMS deviation between the bin heights of cumulative histograms of the distribution of the eigenvalue nearest to 1 versus the distribution of the lowest zero. Figure \ref{bsearch} shows what happens to this quantity for four families of elliptic curves as we try a range of values of $A$.  At the optimal value of $A$, the inverse cubic model is quantitatively better (by between 5.7\% and 12.1\% in the four families we tested) than the excised model. 

The vertical lines on the plots in Figure \ref{bsearch} will be relevant in the next section. 

\subsubsection{Interpreting the optimal value of $A$} \label{sect:A}

\begin{figure}[h!]
\centering
\caption{Distributions of central values of even quadratic twists ($0 < d < 400000$, $d$ prime) of the $L$-function associated with $E11.a3$ and of central values of characteristic polynomials from the excised model.}
\label{values_excised}
\includegraphics[width=10cm]{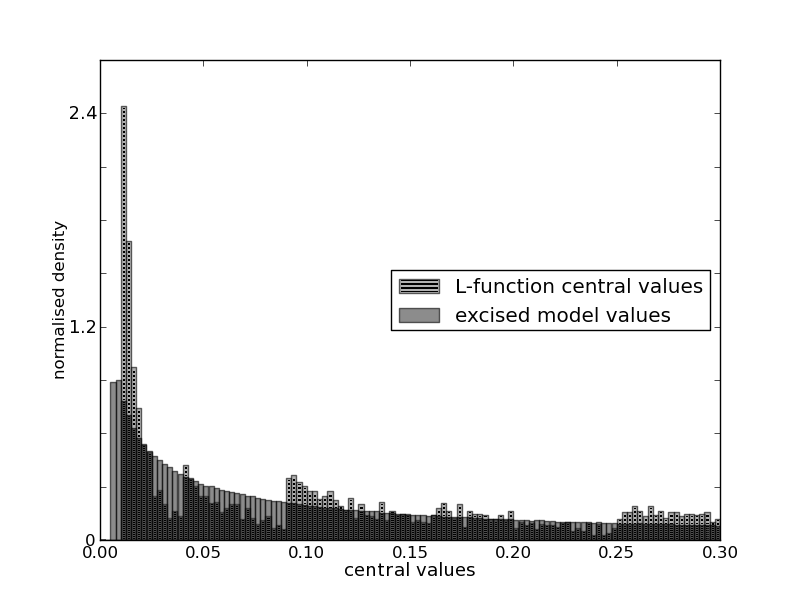}
\end{figure}

Qualitatively, the effect of varying $A$ is to adjust the proportion of matrices in our ensemble with $\Lambda_B(1)< 4\kappa_E X^{-1/2}$.  We would like to use this fact to understand the unexpectedly low cutoff $\delta \kappa_E X^{-1/2}$ in the original excised model.  Consider the first peak in Figure \ref{values_excised}. From that picture, having a cutoff smaller than $\kappa_E X^{-1/2}$ appears to have given a model for the values with a similar proportion of values below $4 \kappa_E X^{-1/2}$ as the number theory data, albeit with a different shape. To quantify this, the full $SO(24)$ central value distribution has around 22.5\% of its `mass' below $4 \kappa_E X^{-1/2}$ (note that we are not scaling by arithmetic factors here since (\ref{eq:PEPO}) is only valid near the origin), while the same quantity for the distribution of $L$-function values has around 13.4 \%. The excised model with the surprisingly small cutoff has 14.7\% in this range.  These values are not identical, but the choice of $4 \kappa_E X^{-1/2}$ as delimiter for ``mass near the origin" of the distribution is somewhat arbitrary.  The numbers 13.4 \% and 14.7\% simply serve to give some indication that the amount of mass in a region close to the origin is important. 

\begin{figure}[h!]
\centering
\caption{Distributions of central values of even quadratic twists ($0<d < 400000$, $d$ prime) of the $L$-function associated with $E11.a3$ and of central values of characteristic polynomials from the inverse cubic model with the parameter $A$ taking the value $A_1$, as described at  (\ref{eq:A1}).}
\label{values_weighted}
\includegraphics[width=10cm]{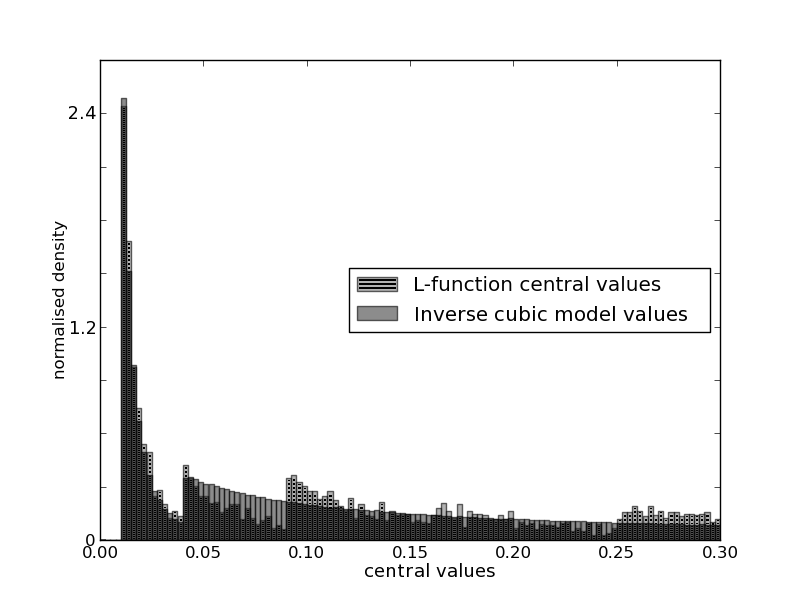}
\end{figure}

Therefore, we propose to calculate the value of the parameter $A$ that ensures that the proportion of matrices  with with $\Lambda_B(1)< 4\kappa_E X^{-1/2}$ exactly matches the proportion of the $L$-function distribution over the same range. We shall call this particular value $A_1$. If we think of the excised model as having unexpected mass below $\kappa_E X^{-1/2}$ in the distribution, then we are accounting for this mass in the new model by moving the cutoff back to its `natural' value and increasing the size of the first peak to compensate. This can be visualised by comparing Figures \ref{values_excised} and \ref{values_weighted}. 

The proportion of mass in the first peak in the distribution of $L$-values is easy to find by counting the number of twisted $L$-functions in our family whose central value lies in this region, and then dividing this quantity by the total number of $L$-functions. The equivalent quantity in the inverse cubic ensemble depends on $A$. As such, we may calculate $A_1$ by equating the proportion of twists for which $\kappa_E X^{-1/2} < L_E(s,\chi_d) < 4\kappa_E X^{-1/2}$ (which we shall denote $S_L$) with the equivalent proportion of our matrix ensemble:
\begin{equation}\label{match}
S_L \coloneqq \frac{1}{X^*} \ \# \{ \kappa_E X^{-1/2} < L_E(s,\chi_d) < 4\kappa_E X^{-1/2} \} \  = \int _{\kappa_E X^{-1/2}} ^{4 \kappa_E X^{-1/2}} P_{\mathrm{ic}}(N,y) \mathrm{d}y.
\end{equation}
The left hand side of (\ref{match}) is computed numerically. We may calculate the right hand side by using (\ref{P_ic}) and (\ref{eq:N_W}):
\begin{align}
\int _{\kappa_E X^{-1/2}} ^{4 \kappa_E X^{-1/2}} P_{\mathrm{ic}}(N,y)\mathrm{d}y  &= N_W \int _{\kappa_E X^{-1/2}} ^{4 \kappa_E X^{-1/2}}  A_1y^{-3}\mathrm{d}y\nonumber \\
&=  \frac{\int _{\kappa_E X^{-1/2}} ^{4 \kappa_E X^{-1/2}}  A_1y^{-3}\mathrm{d}y} {\int _{0} ^{\infty }  W(N,y)P_{O}(N,y) \mathrm{d}y } \nonumber \\ 
&= \frac{\int _{\kappa_E X^{-1/2}} ^{4 \kappa_E X^{-1/2}}  A_1y^{-3}\mathrm{d}y} {\int _{\kappa_E X^{-1/2}} ^{4 \kappa_E X^{-1/2}}  A_1y^{-3}\mathrm{d}y + \int _{4 \kappa_E X^{-1/2}} ^{\infty}  P_O(N,y)\mathrm{d}y }\nonumber \\
&= \frac  {A_1\frac{15X}{32\kappa_E^2}}    {A_1\frac{15X}{32\kappa_E^2} +\int _{4 \kappa_E X^{-1/2}} ^{\infty}  P_O(N,y)\mathrm{d}y  }.
\end{align}
Now $ \int _{4 \kappa_E X^{-1/2}} ^{\infty}  P_O(N,y)\mathrm{d}y  $ may be calculated numerically by  generating $SO(2N)$ matrices chosen with respect to Haar measure and finding the proportion for which $\Lambda_B(1) > 4 \kappa_E X^{-1/2}$. We shall refer to this quantity as $S_H$. We can now rearrange to find $A_1$ in terms of $S_L$ and $S_H$:
\begin{equation}\label{eq:A1}
A_1 = \frac{32S_L S_H \kappa_E^2}{15X(1-S_L)}.
\end{equation}
 For twists up to $d = 308317$ of the $L$-function associated with $E11.a3$, this method gives $A_1 \approx 2.580\times 10^{-5}$.

As can be seen in Figure \ref{mod_weight_zeros_eigs}, we find that this value of $A$ does indeed give us a matrix ensemble whose first eigenvalues are a good model for the lowest zeros of rank 0 quadratic twists. This is illustrated by marking the position of $A_1$ on Figure \ref{bsearch} where it lies extremely close to the optimum.

Note that in calculating $A_1$ we must necessarily generate matrices with integral dimension.  In order to minimise the effect of having to round the value of $N$ given by  (\ref{eq:equatingdensities}) to the nearest integer when generating the matrices, for Figure \ref{bsearch} we choose $X$ such that  (\ref{eq:equatingdensities}) produces a value of $N$ that is as close to 12 as possible. For $M = 11$, this gives $X = 308317$, and for $M = 19$, we take $X = 234605$.

\subsubsection{Consequences of the inverse cubic model}\label{sect:confessions}

As can be seen in Figure \ref{bsearch}, the parameter value $A_1$ is very good at predicting the optimal value of the parameter $A$ for the four families that we tested. This supports the hypothesis that having approximately the correct mass in this initial portion of the value distribution is a critical feature in a good model and is the reason that, despite their differences, the inverse cubic model and the excised model show similar behaviour in the statistics of the first eigenvalue. Thus, the unexpectedly small cutoff  $\delta \kappa_E X^{-1/2}$ in the original excised model compensates for the differing shapes of the central value distributions near the origin for the twisted $L$-functions and for the excised $SO(2N)$ ensemble.


As we have already noted, we have chosen to quantify the proportion of small values with the mass in the first peak between $\kappa_E X^{-1/2}$ and $4\kappa_E X^{-1/2}$, but we can see from Figure \ref{values_weighted} that there is more structure in subsequent peaks of the distribution of $L$-values.  This must have some effect on the distribution of first zeros and so may explain why the value $A_1$ does not exactly lie at the minimum in every family. 

It should be stressed that due to its complexity, the inverse cubic model is far less convenient for actually working with than the simpler original excised model. We believe that the present work supports the idea that for modelling the zero statistics of elliptic curve $L$-functions far from the Katz-Sarnak limit, the excised model is indeed a useful tool as an approximation to the more detailed but cumbersome inverse cubic model. 

\subsection{A different approach - the two parameter model}

Since allowing the first cutoff to vary from the natural value of $\kappa_E X^{-1/2}$ proved to be useful in constructing the original excised model, a possible criticism of the inverse cubic model described in Section \ref{sect:modweightmod} is that we have fixed the position of the two `excision points'  (the points of discontinuity in Figure \ref{weighty}).  It is natural to ask whether or not we should consider moving the positions of these excision points in a modified excised ensemble, instead of scaling the height of the first spike. 

\begin{figure}[h!]
\centering
\caption{An example of the distribution of $\Lambda_B(1)$ values from the two-parameter model, with $\chi_1 = 0.026$ and $\chi_2 = 0.052$. Note that the height of the first peak will follow the envelope of $P_O(N,y)$ as we vary $\chi_1$.}
\label{two_param_sketch}
\includegraphics[width=10cm]{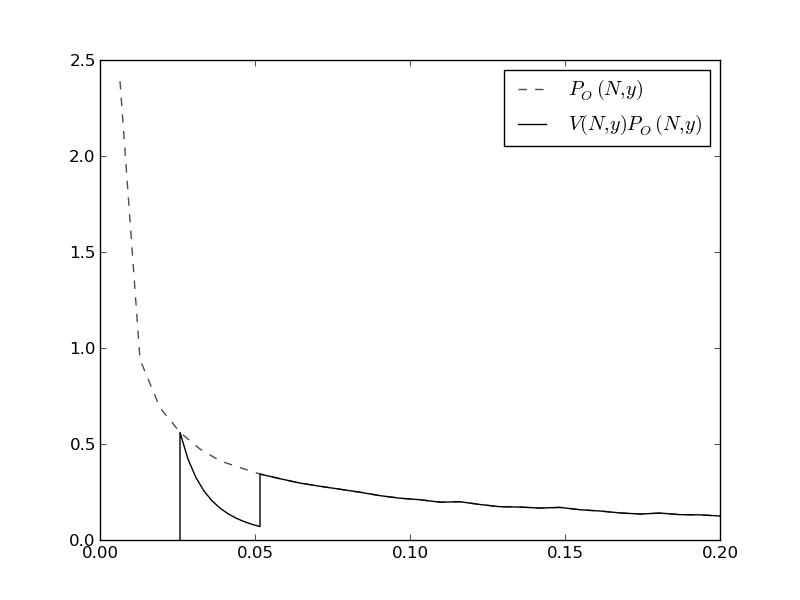}
\end{figure}

We once again consider an ensemble with a $\Lambda_B(1)$ value distribution displaying a sharp turn on followed by an inverse cubic decay, but we shall now fix the height of the first peak in the distribution of our values. We shall implement this in a similar manner to Section \ref{sect:modweightmod}  by defining a weight $V_{\chi_1,\chi_2}(N,y)$ (from here on we will drop the subscripts on $V$). This weight has the property that if $\chi_1$ is the value of the first excision point, then $V(N,\chi_1) = 1$. (Recall that in the inverse cubic model we allowed this height to be greater than 1.)  Between $\chi_1$ and the second excision point $\chi_2$, $V(N,y)P_O(N,y)$ follows an inverse cubic curve as before. For a sketch of $V(N,y)P_O(N,y)$ see Figure \ref{two_param_sketch}.  Note that for values greater than $\chi_2$, $V(N,y)$ takes the value 1, as in Figure \ref{weighty}. We shall refer to this as the `two parameter model', where the two parameters are $\chi_1$ and $\chi_2$. 

\begin{figure}[h!]
\centering
\caption{The RMS deviation of the two parameter model as a function of the two excision parameters for four families.  The `X' marks the minimum deviation. The family of positive twists of $E11.a3$ had a cutoff (defined at (\ref{eq:RMTcutoff})) of $c\exp(-N/2)=0.0048$ in the original excised model with $0<d\leq 400 \;000$. The family of negative twists of $E11.a3$ had a cutoff of $c\exp(-N/2)=0.0023$ in the original excised model with $-300 \;000\leq d <0$. The family of positive twists of  $E19.a2$ had a cutoff of $c\exp(-N/2)=0.0027$ in the original excised model with $0<d\leq 480 \;000$. The family of negative twists of $E19.a2$ had a cutoff of $c\exp(-N/2)=0.0034$ in the original excised model and $-300\;000\leq d<0$. }
\label{rainbows}
\includegraphics[width=7cm]{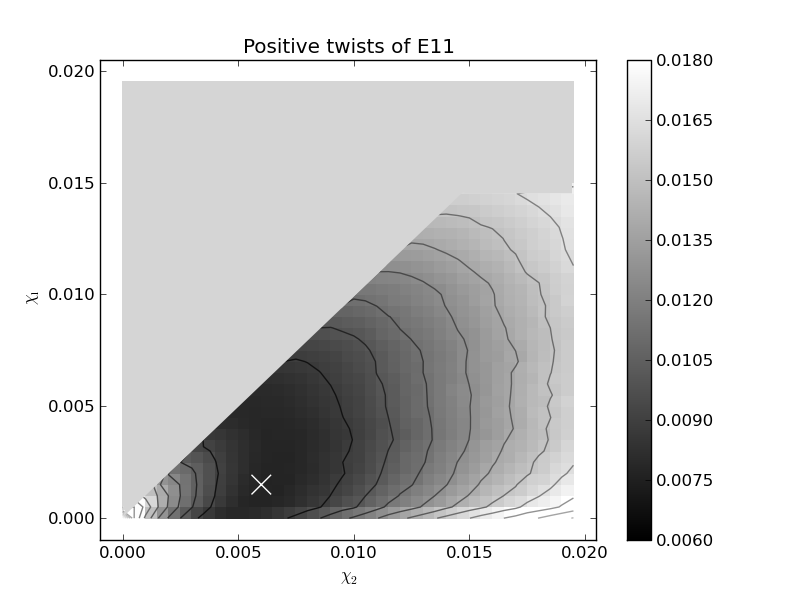}
\includegraphics[width=7cm]{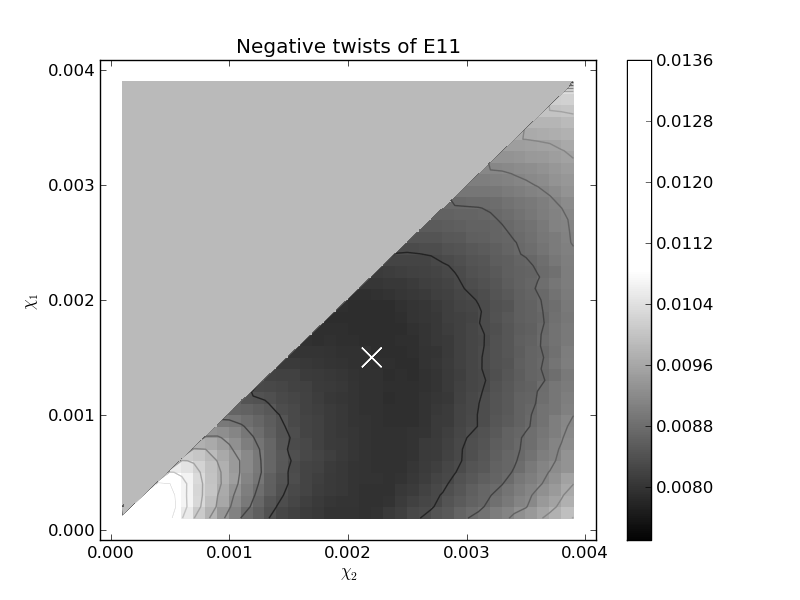}
\includegraphics[width=7cm]{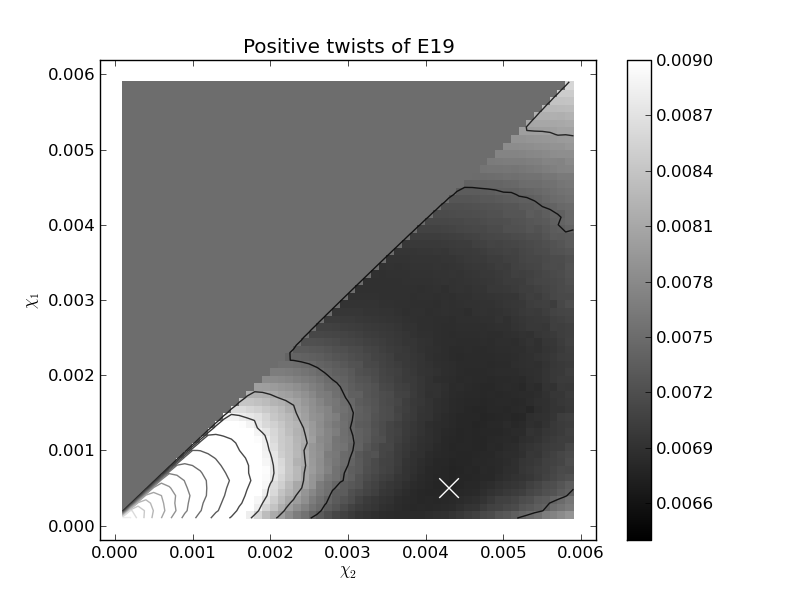}
\includegraphics[width=7cm]{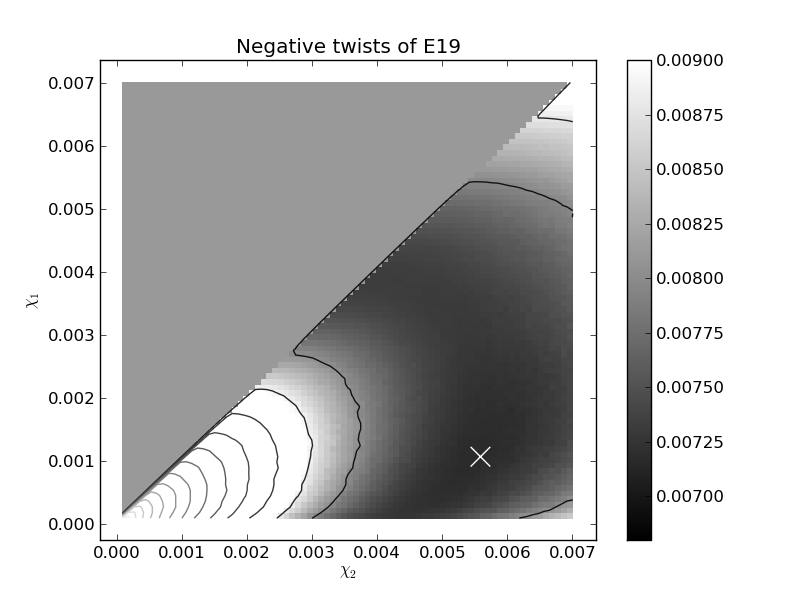}

\end{figure}

As before, we consider the RMS difference between the cumulative distribution of the eigenvalue nearest 1 in the two-parameter model and the cumulative distribution of the lowest of the $L$-function zeros. We would like to know if there is a pair of values for the two parameters $\chi_1$ and $\chi_2$ which would give us a significant improvement on the original excised model. This is an optimisation problem in two dimensional phase space, and it is computationally viable to find the optimal values of the two parameters numerically. This was done for both positive and negative twists of $E11.a3$ and $E19.a2$ to give us the optimum two parameter model for four sets of data.  The plots in Figure \ref{rainbows} show this quantity as a function of the two excision parameters.

\subsubsection{Discussion of the two-parameter model}

 Note that when the values of $\chi_1$ and $\chi_2$ coincide, on the diagonal of the plots, this model reverts to the original excised model.  For all the families considered, the optimum value (marked by a cross on the plots) is in fact off the $\chi_1 = \chi_2$ line. That is to say, there are parameter values that improve on the excised model.  However, the percentage improvements (all around 0.2\% better than the original excised model) are so small as to be inconclusive when compared to the inverse cubic model, which produced much larger percentage improvements over the original excised model (5.7\% to 12.1\% for the four families we tested). Thus we conclude that the real insight is to be gained from the inverse cubic ensemble.
 
 Interestingly, although the two-parameter model does not significantly improve on the original excised model, it appears to be consistent with the idea that it is the proportion of small values of $\Lambda_B(1)$ in the ensemble that is a good indicator of how accurately the ensemble models the statistics of the zeros.  Consider again the diagonal of the plots in Figure \ref{rainbows}, which corresponds to models where the distribution of $\Lambda_B(1)$ has only a single discontinuity.  At the origin, this discontinuity is at zero and this gives the full $SO(2N)$ ensemble.  As we move along the diagonal, this discontinuity in the value distribution moves away from the origin.  The darkest region on the diagonal of the plots in Figure \ref{rainbows} illustrates the optimal position for this discontinuity and corresponds to the minimum in the corresponding plot in Figure \ref{fig:RMSdeviations}.  The excised model cutoffs (defined at (\ref{eq:RMTcutoff})) are listed at Figure \ref{rainbows} for comparison. The $\chi_2$-axis represents a model with the first excision point, $\chi_1$, very close to the origin.  When $\chi_2$ is also small there is a larger proportion of small central values of $\Lambda_B(1)$  than we find in the distribution of central $L$-values, as is the case for $SO(2N)$.  Moving from left to right in the plots in Figure \ref{rainbows} this proportion will tend to reduce, resulting in a growing repulsion of the lowest eigenvalue.  In correlation with the decreasing mass, the RMS deviation reaches a minimum and then increases again as we move from left to right.  Along a similar line of reasoning, the white `X' on the plots indicates the minimum RMS deviation between the model and the distribution of the first zero of our family of $L$-functions, and it generally lies to the right and below the minimum on the diagonal, implying that the second excision point has moved to the right, reducing the mass near the origin, and to compensate the first excision point has moved towards the origin so a little more mass is contributed from the first peak.  These are speculations, as we did not pursue this model in further detail, but it appears to be consistent with the hypothesis inspired by the inverse cubic model that the proportion of small values is critical to a  good model.

\clearpage


\end{document}